\documentclass[11pt,a4paper]{article}
\usepackage[utf8]{inputenc}
\usepackage{graphicx}
\usepackage{authblk}
\usepackage[version=3]{mhchem}
\usepackage[left=3cm,right=3cm,top=3cm,bottom=3cm]{geometry}

\begin{document}
\title{Kinetics of Electron-Beam Dispersion of Fullerite \ce{C60}}
\author[1]{Ihar Razanau \thanks {ir23.by@gmail.com}}
\author[2]{Tetsu Mieno}
\author[1]{Victor Kazachenko}
\affil[1]{Department of Materials Science and Technology, Belarusian State University of Transport, Gomel, 246653, Belarus}
\affil[2]{Department of Physics, Shizuoka University, Shizuoka, 422-8529, Japan}
\renewcommand\Authands{ and }
\date{13 December 2011}
\maketitle
\begin{abstract}
		Electron-beam dispersion of pressed fullerite \ce{C60} targets in vacuum leads to the deposition of thin films containing polymeric forms of \ce{C60}. The aim of the present report is to analyze physical-chemical processes in the fullerite target during its electron-beam dispersion through the analysis of the kinetics of the radiation temperature of the target surface, the coating growth rate and the density of negative current on the substrate. It was shown that the induction stage of the process is determined by the negative charging and radiation-induced modification and heating of the target.  The transitional stage is characterized by nonstationary sublimation of the target material through the pores in the modified surface layer and release of the accumulated negative charge. Stabilization of the process parameters owing to the convection cooling of the target by the sublimation products and the decrease in the pressure inside the microcavities beneath the pores leads to a quasi-stationary stage of target sublimation and deposition of a coating containing polymeric forms of \ce{C60}.\\
		
\end{abstract}

\begin{small}
\begin{center}
Published in Nucl. Instrum. Methods Phys. Res., Sect. B 280 (2012) 117–122\\
DOI: 10.1016/j.nimb.2012.03.015\\
\end{center}
\end{small}

\section{Introduction}
Electron beam irradiation of fullerite \ce{C60} can lead to various physical-chemical effects ranging from molecular structure modification to bulk explosive sublimation. The effect depends on the energy of the electrons, the current density of the electron flux, the exposure dose and the conditions of the energy dissipation.\\

Polymerization of thin \ce{C60} films was observed upon irradiation by electrons emitted from a scanning tunneling microscope (STM) tip with energy of 3--4~eV~\cite{1,2,3,4}. Fullerene fragmentation and formation of coalescent spheroid structures was detected with increase of the energy of the electrons emitted from a STM tip to 20--45~eV~\cite{5} and to 120--310~eV~\cite{6}. The authors suggested that heating effect is negligible under the irradiation conditions used (the exposure dose of up to 7200~nC and the electron flux of $\approx10^{-3}$--$10^{-1}$~nA/nm$^2$) and therefore, it is electron excitation that causes fragmentation and coalescence of \ce{C60} molecules.\\

Polymerization, hydrogenation and consequent fragmentation and graphitization of ultrathin \ce{C60} films grown on hydrogen passivated Si~(100) substrates was observed after irradiation with 0.5~keV and 3.3~keV electrons. The electron beams were produced by an electron gun with a nominal maximum of electron flux density of $\approx64$~mA/cm$^2$~\cite{7}. The films were exposed to the dose of up to $3\cdot10^{-2}$~C/cm$^2$ for the 0.5~keV beam and $9\cdot10^{-1}$ C/cm$^2$ for the 3.3~keV beam. No measurable desorption of the film material was detected under the irradiation conditions used. Thermally reversible polymerization was reported for ultrathin \ce{C60} films on a GaAs~(110) substrate exposed to a defocused 1.5~keV electron beam with the flux density of $\approx1.6$~mA/cm$^2$ and the exposure dose of up to $\approx18$~C/cm$^2$~\cite{1}. The polymerization was described as a nucleation-and-growth process influenced by the irradiation-induced heating. Formation of peanut-shaped \ce{C60} polymers in a 70-nm-thick \ce{C60} film grown on a CsI substrate was observed upon irradiation by a scanning 3~keV electron beam~\cite{8}. The electron flux density used was $\approx0.1$--0.3~mA/cm$^2$. The nonlinear dependence of the reaction rate on the electron dose rate was found. The authors have concluded that a complex reaction between secondary electrons and intermediate products leads to coalesced \ce{C120} dimer formation.\\

A normalized modification parameter $P$ was proposed on the basis of the relative change of the low-energy part of the electron energy loss spectrum of \ce{C60} films under 150--1500~eV electron irradiation. It was found that the modification parameter is a monotonously raising function of the irradiation dose and the incident electron energy. Fast change of the $P$ under small irradiation doses corresponds to \ce{C60} polymerization. Subsequent saturation of the $P$ under high irradiation doses corresponds to \ce{C60} fragmentation and amorphization~\cite{9, 10}. It should be noted that a low-intensity electron beam of the electron spectrometer was used for irradiation. Therefore, the irradiation-induced heating can be considered negligible under the conditions given.\\

A 20~keV tubular electron beam with a current of 50~mA focused into a spot of 3~mm in diameter (the current density of $\approx700$~mA/cm$^2$) was used for explosive sublimation of a fullerene target within less than 1~s and simultaneous deposition of a nonpolymerized fullerene film onto a substrate~\cite{11}. In another report~\cite{12}, a 20~keV electron irradiation was used to decrease significantly the dissolution rate of a \ce{C60} film owing to fullerene polymerization and subsequent graphitization. The film was exposed to the irradiation dose of up to 0.024~C/cm$^2$. As no detectable film desorption was observed, we can suppose that the electron beam used had a considerably lower current density in comparison with the previous case.\\

We have previously used the method of electron-beam dispersion (EBD) for synthesizing thin films on the basis of fullerene \ce{C60}. In the EBD method, an electron beam of 1--1.5~keV energy and 30--50~mA current is used to heat and sublimate the target made of pristine material in high vacuum. The coating is deposited onto a substrate from the volatile sublimation products typically containing neutral and excited molecules, ions and radicals. It was shown that the coatings produced by EBD of fullerite \ce{C60} contain \ce{C60} polymeric forms~\cite{13}. In this report, we concentrate on the analysis of the kinetics of fullerite \ce{C60} EBD. On the basis of the obtained experimental data and simple physical estimates, physical-chemical processes in the fullerite target during its EBD are speculated about.

\section{Experimental details}
Experimental setup for studying the kinetics of fullerite EBD is shown in Fig.~1. 80~mg of fullerite \ce{C60} powder (purity $> 99.5$~wt.\%) was pressed into a disc of 5~mm in diameter and of $\approx2.5$~mm thickness under the pressure of $\approx25$~MPa in 30~s. The target was placed on a copper heating table in a stainless steel vacuum chamber. The vacuum chamber was evacuated by a turbomolecular pump and kept under the pressure of $1\cdot10^{-2}$~Pa. The fullerite target was then heated up to 473~K and annealed at this temperature for 1~hour. Vacuum annealing of the fullerite target was introduced to decrease the amount of oxygen absorbed by fullerite.\\

\begin{figure}[h!]
 \begin{center}
 		\includegraphics[scale=1]{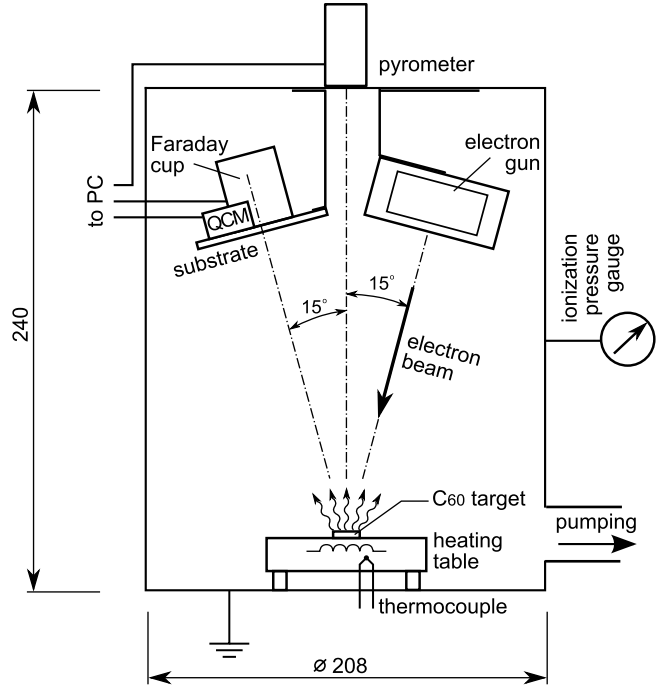}
		\caption{Experimental setup for studying the kinetics of electron-beam dispersion of fullerite \ce{C60}.}
\end{center} 
\end{figure}

Electron-beam dispersion of the target was enabled by an electron gun with a directly heated wire cathode of tungsten. Electron beams with the energy of 1--1.5~keV and the current of up to 50~mA were used for EBD. However, we have found that the electron beam with the energy of 1.2~keV and the current of 30~mA enables the most stable EBD of fullerite \ce{C60} targets under the experimental conditions given. Hereafter, we will describe and analyze EBD of fullerite \ce{C60} by the 1.2~keV, 30~mA electron beam. The electron beam spot was larger than the target thus enabling its uniform irradiation. The current density of the electron beam irradiating the surface of the target was about 10~mA/cm$^2$ according to the measurement by a Faraday cup. It should be noted that the measurement of the current density was performed by replacing the target holder with the Faraday cup and thus, it does not consider possible target charging upon irradiation.\\

The temperature of the target surface was registered by an IMPAC IPE 140 radiation pyrometer with a PbSe infrared sensor (spectral range of 3--5~$\mu$m) through a vacuum window made of a KRS-5 crystal. The pyrometer enabled measurement of the surface temperature of the target in a spot of 0.7~mm in diameter. The emissivity coefficient $\epsilon\approx0.85$ characteristic of graphite in the spectral range of 3--5~$\mu$m was used for temperature measurement. The transmission characteristics of the KRS-5 window were also taken into account.\\

The emissivity coefficient depends on the surface condition of the material and the measuring temperature. For fullerene \ce{C60} below 1000~K, the heat radiation is mainly emitted by infrared-active vibrations~\cite{14}. As \ce{C60} does not have infrared-active absorption bands in the spectral range used by the pyrometer, its emissivity is thought to be significantly lower than that of graphite. Hence, in the beginning of the EBD process the real temperature of the target surface is higher than the radiation temperature registered by the pyrometer. During the EBD, the temperature of the target surface increases by several hundred kelvins and the surface of the target is modified by the electron irradiation. It will be shown later that the structure of the target surface changes from pristine \ce{C60} to disordered carbon upon EBD. Thus, the difference between the real temperature and the radiation temperature is thought to decrease with time of electron-beam irradiation.\\

The effective mass thickness of the coating deposited on the substrate was monitored by a quartz crystal microbalance~(QCM). To recalculate the obtained data into the geometrical thickness, the mass density of the deposited material was chosen to be equal to the mass density of fullerite \ce{C60}~(1720~kg/m$^3$) as no micro-size pores or defects were found in the deposited films using the atomic-force microscopy measurement~\cite{13}.\\

A Faraday cup was placed into the substrate holder together with the QCM sensor. The Faraday cup registered the current of scattered and true secondary electrons and ions from the target and the copper heating table used as the target holder.\\

To summarize, the kinetics of the radiation temperature of the target surface, thickness of the coating on the substrate and current irradiating the substrate during the EBD process was registered. The first-order derivatives of these three parameters representing growth rate of the coating, heating rate of the target surface and the rate of the substrate irradiation change were also analyzed.\\

A Jasco NR~1800 spectrometer with a cooled CCD detector was used for Raman spectroscopy measurement of the fullerite target before and after the EBD process. The 488~nm line of an argon laser was used as the excitation source.

\section{Results and discussion}
The whole process of fullerite \ce{C60} EBD can be divided into an initial induction stage, a relatively short transitional stage and a quasi-stationary stage of target sublimation and coating deposition (Fig.~2). The induction stage of fullerite \ce{C60} EBD is characterized by target negative charging, target heating and irradiation-induced modification of the target surface layer. During the induction stage, the electron beam irradiates the target yet the sublimation of the target material is negligible. At the beginning of the electron-beam irradiation, the density of the negative current onto the substrate decreases from initial $\approx1.1$~mA/cm$^2$ to a value of $\approx0.7$~mA/cm$^2$ within few seconds (Fig.~3a).\\

\begin{figure}[h!]
 \begin{center}
 		\includegraphics[scale=1]{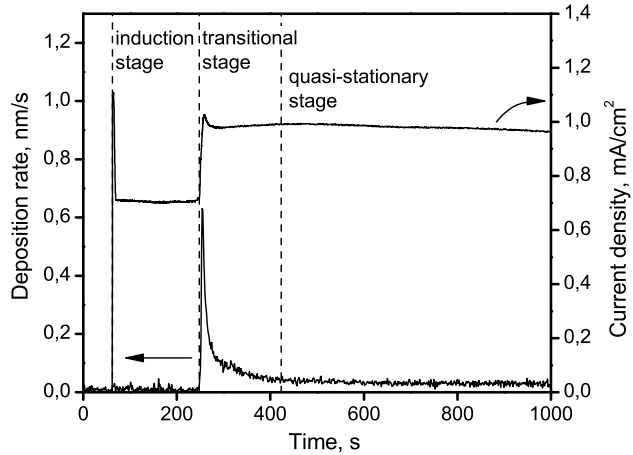}
		\caption{Kinetics of the coating deposition rate and the negative current density on the substrate during the EBD of fullerite \ce{C60}.}
\end{center} 
\end{figure}

In our opinion, the negative charging of the target is caused by the following reasons. The coefficient of secondary electron emission (SEE) is below~1 for carbon-only materials in the electron energy range of 1000--1500~eV~\cite{15}. Moreover, fullerite~\ce{C60} has low conductivity at room temperature. We will discuss later how the increase of the fullerite conductivity upon heating can significantly influence the target charging.\\

During the induction stage, the substrate is irradiated by a flow of secondary and scattered primary electrons from the target and the target holder.  A slight change of the geometry of the wire cathode of the electron gun upon its heating causes relatively small decrease of the primary electron beam current and hence, it can also cause the decrease of the current density on the substrate. However, in our opinion, the decrease observed is mainly caused by the charging of the target.\\

The electric field of the negatively charged surface of the fullerite target is thought to suppress and scatter secondary electrons emitted from the target holder surface around the target. We have measured the energy distribution of the electron flow on the substrate under the experimental conditions given without the fullerite target. The measurement was conducted using a Faraday cup with a copper mesh in front of it. A decelerating negative electrostatic potential from 0 to 300~V was applied to the mesh. The measurement showed that secondary electrons with the energy of less than 50~eV constitute $\approx37$\% of the whole electron flow under the experimental conditions given. The value obtained is quite close to the decrease of the current density on the substrate observed during the EBD of fullerite.\\

The analysis of the kinetics of the current density on the substrate showed that its decrease upon the beginning of the electron-beam irradiation indeed can be fitted by a sum of two exponential decay functions (Fig.~3a). A function with the time constant of about 8~s and the amplitude of about 0.15~mA/cm$^2$, which we connect to the particularities of the electron gun operation, and a function with smaller time constant of about 1~s and larger amplitude of about 0.42~mA/cm$^2$. As explained above, we connect the second decay function to the suppression and scattering of SEE from the target holder.\\

\begin{figure}[h!]
 \begin{center}
 		\includegraphics[scale=1]{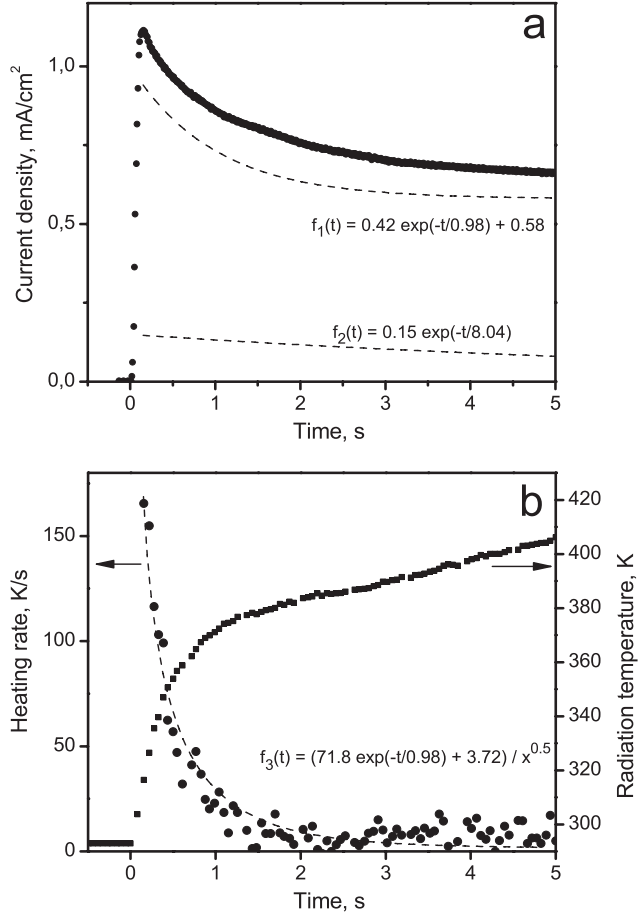}
		\caption{Kinetics of the negative current density on the substrate~(a), the radiation temperature, and the heating rate of the target surface~(b) with respective fitting functions (dashed lines) during the negative charging of the fullerite target under the electron-beam irradiation.}
\end{center} 
\end{figure}

During the induction stage, the primary electron beam is decelerated and scattered by the electrostatic potential of the target surface. Consequently, the power density of the electron beam on the target and its heating rate also decrease during the first few seconds of fullerite EBD~(Fig.~3b). Fullerite~\ce{C60} has relatively low thermal conductivity (0.4~W$\cdot$m$^{-1}\cdot$K$^{-1}$ for \ce{C60} crystals~\cite{16}). Thus, the fullerite target can be considered as a semi-infinite body during the first seconds of electron irradiation until the heating front reaches the bottom of the target. For a semi-infinite body, the heating rate of the surface is proportional to $q/\sqrt{t}$, where $q$ is the power density of the heat flux and $t$ is the time. However, the negative electrostatic potential of the target is thought to decelerate and defocus the primary electron beam. Therefore, we have used the $f(t)=(C_1\cdot\exp(-t/\tau)+C_2)/\sqrt{t} $ formula with an exponential decay function instead of the constant $q$ to fit the experimental data~(Fig.~3b). The time constant $\tau\approx0.88$~s, obtained by the fitting, is quite close to the time constant of the decay of the current density on the substrate ($\tau\approx0.98$~s). This agreement proves that both decay processes are caused by the same reason, negative charging of the target surface.\\

After the negative charging of the target is completed, the rest of the induction period is characterized by further slow growth of the target temperature and irradiation-induced modification of a thin surface layer of the target. The penetration depth of the primary electrons into the fullerite target can be estimated by using a formula proposed by Kanaya and Okayama~\cite{17}. For 1.2~keV electrons it is $\approx53$~nm. It should be noted that the primary electrons are decelerated by the target negative charge and thus, the calculated value is an upper estimate.\\

The electron-induced modification of the surface target layer goes through polymerization to fragmentation and amorphization upon the increase of the exposure dose~\cite{9, 10}. However, fullerene polymers are metastable and undergo spontaneous dissociation to monomer in the temperature range of 400--600~K~\cite{18,19,20,21}. Therefore, heating and polymerization/fragmentation of the target material are two interconnected processes in the surface layer of the target whereas in the deeper layers, heat transfer prevails.\\

The duration of the induction period and the scenario of the sublimation start depend strongly on the thermal physical parameters of the target and the thermal contact between the target and the target holder. Under the experimental conditions used, the target sublimation starts after approximately 3~minutes of electron-beam irradiation and goes through pores in the modified surface layer. The porous surface layer can be seen as a thin flaky film having metallic luster on the fullerite targets after the EBD process~(Fig.~4b). On the other hand, preparation of thinner and denser targets of higher thermal conductance results in longer induction times and the start of the sublimation through an explosive take-off of the flakes of the dense modified barrier layer by the overheated subsurface fullerenes. In the areas, where the surface film has been taken away, the direct surface sublimation prevents the recurring formation of the irradiation-induced barrier layer. These areas can be seen as pristine-black eroded spots on the fullerite targets after the EBD~(Fig.~4c,d).\\

\begin{figure}[h!]
 \begin{center}
 		\includegraphics[scale=1]{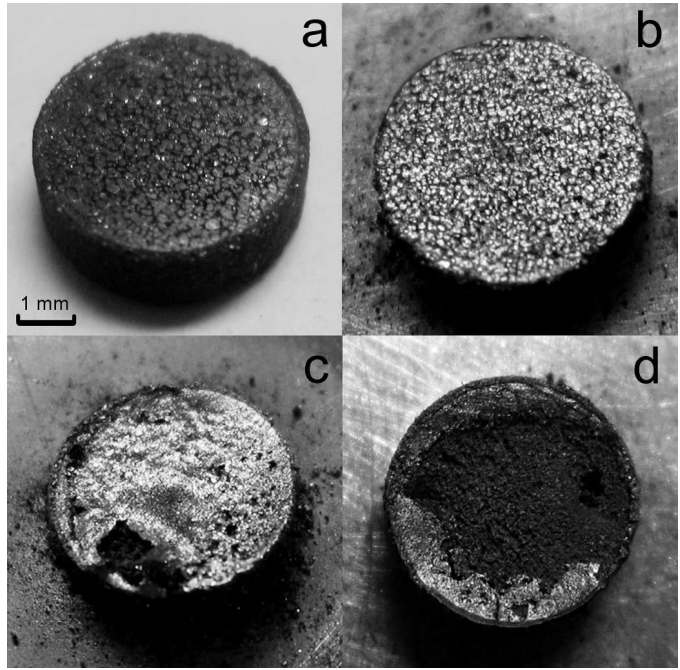}
		\caption{Photographs of the fullerite \ce{C60} targets before~(a) and after the process of EBD~(b, c, d); in the case of target~(b), the sublimation went through the pores in the surface layer, whereas in the case of targets~(c) and~(d), parts of the surface layer were explosively taken off during the start of sublimation.}
\end{center} 
\end{figure}

The material of the thin modified surface layer corresponds to amorphous carbon with inclusions of fullerene \ce{C60} polymers. It is insoluble in toluene. It was purified by a repeated toluene extraction of unreacted \ce{C60}. The Raman spectrum of the material drastically differs from the spectrum of pristine \ce{C60}~(Fig.~5). Two broad and intensive peaks at around 1357 and 1585~cm$^{-1}$ are characteristic of disordered carbon structures with a mixture of sp$^2$ and sp$^3$ carbon sites and known as D and G~peak, respectively~\cite{22}. Among the less intensive peaks, the peak at 1465~cm$^{-1}$ can be assigned to \ce{C60} polymers.\\

\begin{figure}[h!]
 \begin{center}
 		\includegraphics[scale=1]{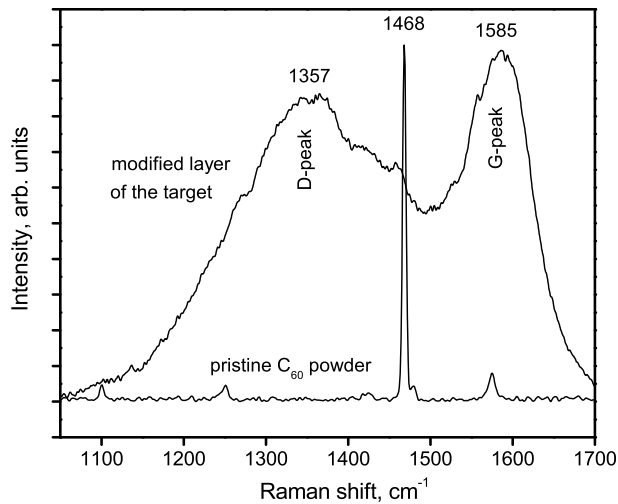}
		\caption{Raman spectra of pristine fullerite \ce{C60} powder and the electron-modified surface layer of fullerite \ce{C60} targets after the EBD process.}
\end{center} 
\end{figure}

The non-stationary coating deposition stage of the fullerite~\ce{C60} EBD is triggered by the release of the accumulated negative charge with the first sublimation products. Consequently, the negative current density onto the substrate increases from $\approx0.7$ to $\approx1.0$~mA/cm$^2$ (Fig.~6b). The observed increase corresponds to the electron flow density of approximately $1.9\cdot10^{15}$~s$^{-1}$cm$^{-2}$, whereas under the experimental conditions given and the stipulation of 100\% accommodation of the depositing molecules, the maximum flow density of depositing fullerene molecules was $\approx1\cdot10^{14}$~s$^{-1}$cm$^{-2}$ ($\approx0.7$~nm/s). Hence, a possible contribution of the ion current is negligible. The recovered value of the current density is close to its value before the negative charging of the target surface ($\approx1.1$~mA/cm$^2$). Thus, the target may be only slightly charged during the following stages of the EBD process.\\

The increase of the power density of the electron beam on the target owing to its discharging enables a fast raise of the target surface temperature. The peak of the heating rate during the first 10~seconds of the sublimation (Fig.~6c) coincides well with the peak of the target discharging rate. The latter can be seen as a peak of the first derivative of the current density on the substrate (Fig.~6b). The fast raise of the target temperature has two main consequences on the EBD process: the increase of the target conductivity and the start of the intensive thermal sublimation.\\

The dominating conductivity mechanism of fullerite \ce{C60} changes with heating. Upon reaching the sublimation temperature, it switches from activated hopping via band tail states to band conduction via delocalized states~\cite{23}. The corresponding increase of the conductivity is thought to enable a leakage current of the target surface charges into the target holder, significant enough to prevent charging of the target during the quasi-stationary deposition stage of EBD. Indeed, we have shown that the content of negative fullerene ions in the sublimation products is negligible during the quasi-stationary stage~\cite{13}. Therefore, the sublimation enables the release of the surface charges only during the transitional non-stationary deposition stage, whereas the leakage current through the target comes into play after the target got sufficiently heated.\\

The start of the intensive thermal sublimation of the target material (Fig.~6a) owing to the increase in the target temperature introduces a new factor into the heat balance of the target. In addition to the heat transfer through the target, the convection heat extraction by the sublimation also starts cooling the target surface. These factors slow down the heating of the target by the electron beam, the power density of which increases owing to the decrease of the target electrostatic potential. As a result, the temperature of the target stabilizes by the end of the transitional stage.\\

\begin{figure}[h!]
 \begin{center}
 		\includegraphics[scale=1]{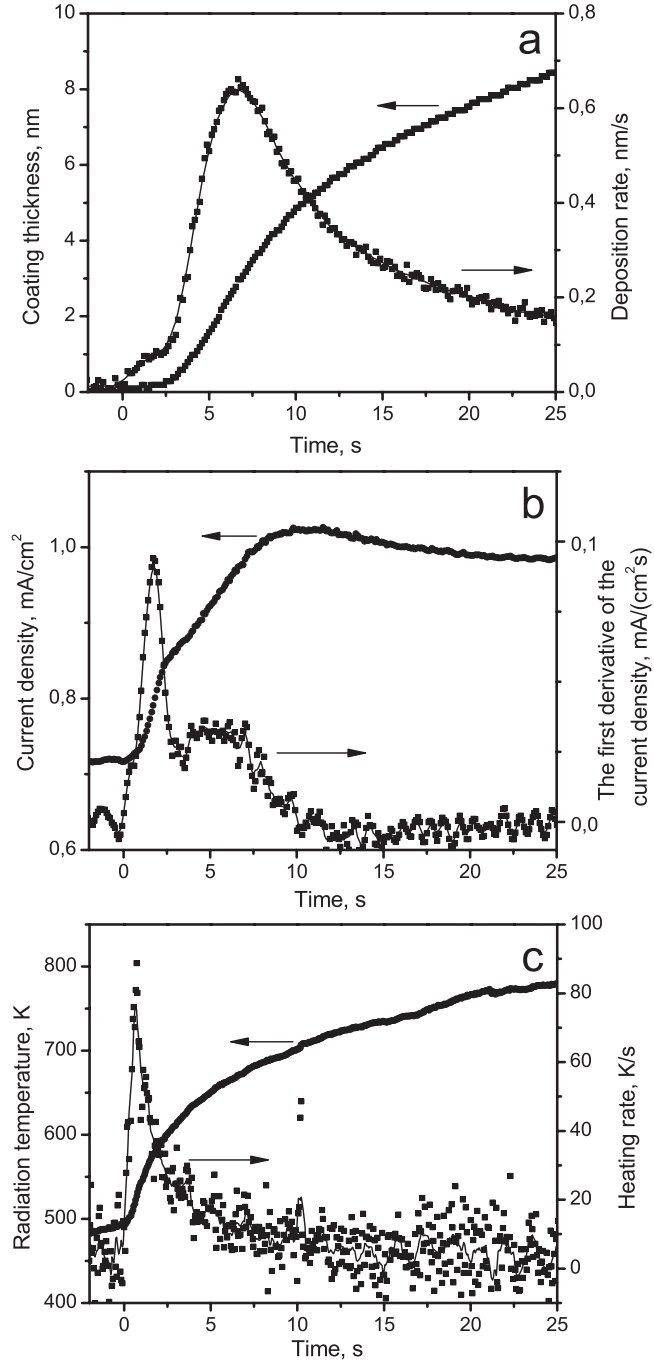}
		\caption{Kinetics of the coating deposition~(a), negative current density on the substrate~(b) and radiation temperature of the target surface~(c) with the respective first derivatives during the transitional stage of non-stationary sublimation (the time is set to~0 at the moment of the coating deposition beginning).}
\end{center} 
\end{figure}

As the sublimation goes through the pores in the amorphous carbon surface layer, microcavities appear and grow under the pores. Indeed, in the case of the direct sublimation from the surface, the sublimation rate is proportional to the target surface temperature. With the stabilization of the temperature, the sublimation rate is also thought to stabilize and not to change significantly until the target is used up. Such sublimation kinetics is observed when parts of the modified surface layer had been taken off during the sublimation start. In the case of the sublimation only through the pores, the sublimation rate from the whole target is proportional to the vapor pressure inside the microcavities. At the constant temperature of the microcavity walls, the pressure inside it is proportional to the sublimation area of the walls and inversely proportional to the volume of the microcavity. As the volume of the microcavity increases faster than its inner surface, we observe a peak on the graph of the sublimation rate rather than step-like increase (Fig.~6a).\\

The modified surface layer of amorphous carbon plays an important role during the stage of the quasi-stationary coating deposition as well. The layer is formed during the induction stage by the primary electrons decelerated in the electrostatic field of the charged target. During the coating deposition stage, the target may be only slightly charged and the primary electrons penetrate through the modified surface layer into the microcavities under it. For electrons, the maximum of the ionization cross section of \ce{C60} lies at around 80--100~eV energy~\cite{24}. Hence, the ionization of \ce{C60} vapor is facilitated inside the microcavities as compared to the direct sublimation from the surface. Indeed, we have previously shown that during the quasi-stationary deposition, the products of fullerite \ce{C60} EBD contain few mass percent of positive fullerene ions with the energy of up to $\approx7$~eV~\cite{13} that strongly influence the coating structure.\\

We define the end of the transitional stage and the beginning of the quasi-stationary coating deposition stage by the stabilization of the coating growth rate and the current density on the substrate. These process parameters determine the conditions of the coating formation on the substrate and therefore, its structure. The coatings used for structural studies described in~\cite{14} were formed during the quasi-stationary deposition stage. In that case we used a shutter to cover the substrates during the first two process stages.\\

The quasi-stationary deposition stage is characterized by the current density on the substrate of 0.95--1~mA/cm$^2$ and the coating deposition rate of 0.01--0.05~nm/s. It should be mentioned that the deposition rate slowly decreases, however within the time limits necessary for depositing a 100--200-nm-thick coating on the substrate, the decrease is negligible. The deposition of the \ce{C60} EBD products onto the substrate under the conditions described results in the formation of a thin film containing a mixture of fullerene \ce{C60} and its polymeric forms.

\section{Conclusions}
Experimental data on the target surface temperature, the density of the flow of charged particles on the substrate, the deposition rate of the \ce{C60} coating, the structure of the modified surface layer of the target have been analyzed. It was shown that the process of irradiation of pressed fullerite \ce{C60} targets with an electron beam with the energy of 1.2~keV and the current density of $\approx10$~mA/cm$^2$ can be considered as a process consisting of several stages.\\

The first, induction stage (the duration of approximately 3~minutes) is characterized by the negative charging of the target and decrease in the current density of the electron beam on the target, heating and radiation-induced modification of the surface layer.\\

The second, transitional stage (the duration of approximately 3~minutes) begins with the loss of the target mass with sublimation of fullerene molecules through the modified surface layer leading to the release of the negative charge and increase in the power density of the electron beam on the target. Sublimation through the pores in the surface layer results in the formation of microcavities under the surface. As a result of the increased power density of the electron beam on the target, the target temperature and consequently, the target conductance and the sublimation rate also increase. Convection cooling by sublimation and the decrease in the pressure inside the microcavities leads to the stabilization of the process.\\

During the third, quasi-stationary stage, the coating is deposited onto the substrate at a rate of 0.01--0.05~nm/s under the electron and ion bombardment leading to the formation of polymerized fullerene structures.

\section*{Acknowledgements}
This work was financially supported by the Belarusian Republican Foundation for Fundamental Research and the Shizuoka University, The True-Nano Project.
\bibliographystyle{ieeetr}
\bibliography{razanau_NIMB}
\end{document}